\pdfoutput=1
\documentclass[]{mandm}
\usepackage{array}
\usepackage[utf8]{inputenc}

\usepackage{amsmath}
\usepackage{graphicx}
\usepackage{hyperref}
\usepackage{adjustbox}
\usepackage{subcaption}
\usepackage{float}
\usepackage{placeins}
\usepackage{stfloats}

\usepackage{xcolor} % Only highlighting changes
\usepackage{ulem}   % -------------------------

\begin{document}

\sloppy

\title[Live first moment imaging and ptychography]{Live processing of momentum-resolved STEM data for first moment imaging and ptychography}

\author[Achim Strauch, Dieter Weber et al]{Achim Strauch$^{1,2,*}$,
  Dieter Weber$^{1,*}$,
  Alexander Clausen$^1$,
  Anastasiia Lesnichaia$^1$,
  Arya Bangun$^1$,
  Benjamin M\"arz$^1$,
  Feng Jiao Lyu$^3$,
  Qing Chen$^3$,
  Andreas Rosenauer$^4$,
  Rafal Dunin-Borkowski$^1$,
  and Knut M\"uller-Caspary$^{1,2}$}

\affiliation{
$^1$Ernst Ruska-Centre for Microscopy and Spectroscopy with Electrons, Forschungszentrum J\"ulich, 52425 J\"ulich, Germany\\
$^2$2nd Institute of Physics, RWTH Aachen University, 52074 Aachen, Germany\\
$^3$Key Laboratory for the Physics and Chemistry of Nanodevices, Department of Electronics, Peking University, Beijing 100871, China\\
$^4$Institute for Solid State Physics, Universität Bremen, Otto-Hahn-Allee 1, 28359 Bremen, Germany\\
$^{*}$Achim Strauch and Dieter Weber contributed equally to this work and share first-authorship.\\
  Corresponding Author: Achim Strauch \email{a.strauch@fz-juelich.de}}

\begin{frontmatter}

\maketitle

\begin{abstract}
A reformulated implementation of single-sideband ptychography enables analysis and display of live detector data streams in 4D scanning transmission electron microscopy (STEM) using the LiberTEM open-source platform. This is combined with live first moment and further virtual STEM detector analysis. Processing of both real experimental and simulated data shows the characteristics of this method when data is processed progressively, as opposed to the usual offline processing of a complete dataset. In particular, the single side band method is compared to other techniques such as the enhanced ptychographic engine in order to ascertain its capability for structural imaging at increased specimen thickness. Qualitatively interpretable live results are obtained also if the sample is moved, or magnification is changed during the analysis. This allows live optimization of instrument as well as specimen parameters during the analysis. The methodology is especially expected to improve contrast- and dose-efficient in-situ imaging of weakly scattering specimens, where fast live feedback during the experiment is required.


\noindent(Received 6 May 2021; revised 6 July 2021; accepted 17 July 2021))
\end{abstract}

\end{frontmatter}

\section{Introduction}

The development of ultrafast cameras for transmission electron microscopy (TEM) such as the pnCCD \citep{MULLER20121119,RYLL20160404}, the Medipix3 chip \citep{PLACKETT20130123}, delay-line detectors \citep{MULLER20150817,OELSNER200110} or the EMPAD \citep{TATE201602} enabled the collection of the full diffraction space up to a flexible cut-off spatial frequency at each scan point in scanning TEM (STEM). 
This technology paved the way for momentum-resolved STEM techniques with high samplings in both real and diffraction space, sometimes being referred to as \mbox{4D-STEM}. Acquisitions with a detector frame rate of several kHz are currently achieved by employing these cameras. In particular, the mapping of electric fields and charge densities down to the atomic scale \citep{Muller2014b}, meso-scale strain, and electric field measurements by nano-beam electron diffraction \citep{MULLER20150817}, and, furthermore, electron ptychography \citep{HOPPE1969,HEGERL1970} have been enabled by this dramatic detector speed enhancement \citep{NELLIST199504,Rodenburg1992,Rodenburg1993,JIANG201807,HUMPHRY201201}.

Due to its excellent dose efficiency \citep{Zhou2020} STEM ptychography as a method to retrieve the complex object transmission function has gained increasing interest.
Four-dimensional data sets combining real and diffraction space information have been shown to provide enormous flexibility in post-acquisition processing. For example, ptychography has been demonstrated to be capable of both resolution improvement and aberration correction after the acquisition using computational methods \citep{NELLIST199504,Gao2017}. It achieves a better signal to noise ratio for weak phase objects than annular bright field or differential phase contrast \citep{SEKI2018118} and allows reconstruction at extremely low dose \citep{doi:10.1063/1.5143213}.

The high data rates require an efficient implementation of advanced methods for imaging contrast via post-processing in order to minimise the duration of numerical processing.
Ultimately, computing and software implementation capabilities are desirable that allow for the live reconstruction of the ptychographic phase and amplitude, first moments, electric fields, and charge densities, for example.

During experiments a region of interest is usually selected via imaging employing conventional STEM detectors. Weakly scattering and beam sensitive specimens, where ptychography and first moment imaging \citep{Waddell1979,Muller2014b} can be most advantageous, generate poor contrast in conventional imaging modes and quickly degrade using typical beam currents in conventional STEM \citep{PEET201908}. 4D-STEM analyses are normally applied after data acquisition and transfer to a data processing workstation. It is therefore not possible to be certain that the selected region and microscope settings were appropriate until after successful reconstruction. For that reason, one often acquires a larger number of data sets, which takes considerable storage space in the case of 4D-STEM. In contrast, a fast implementation of the considered computational methods would allow to perform this data evaluation live during the experiment.

In this study, we demonstrate 4D-STEM continuous live scanning with the simultaneous ptychographic single-sideband (SSB) reconstruction combined with bright field, annular dark field, and first moment imaging, including its divergence which is proportional to the charge density in thin specimens. To this end, the ptychographic algorithm was firstly reformulated mathematically. This allows navigation on the sample and change of microscope parameters with a live view of the reconstruction, alongside signals from other 4D-STEM techniques. Secondly, we use SrTiO$_3$ and In$_2$Se$_3$ to demonstrate the live evaluation capability of our approach in experiments by in-situ processing of the data stream of an ultrafast camera. Thirdly, the results obtained live are validated by conventional post-processing. Particular attention is drawn to the capability of SSB to provide reliable structural images, to reconstruction artefacts arising from processing partial scans and to reconstructing non-periodic objects. Moreover, the performances of SSB ptychography and the enhanced ptychographic engine (ePIE) are compared, interestingly pointing towards SSB being significantly more robust against dynamical scattering in terms of qualitative structural imaging. This article closes with a detailed discussion and a summary.

\section{Materials and Methods}
\subsection{Live imaging}
A continuous live scanning display for large-scale 4D-STEM data benefits greatly from a data processing method where smaller portions of input data
are processed independently and merged progressively into the complete result. For virtual detectors and
first moments, often referred to as "centre-of-mass (COM)", this is trivial to achieve since each detector frame, i.e., diffraction pattern, produces an independent entry in the
final data set which allows frames to be processed individually, accumulating results in a buffer. Displaying
the contents of this buffer at regular intervals provides a live-updating view.

In contrast, ptychography generates results by putting detector frames from the entire data set, or at least a local environment, in relation to each other. Consequently, adapting the methodology so as to circumvent the processing of the entire data set at once, thus gradually merging partial results extracted from portions of the input data into the complete result, is necessary.
As previously demonstrated by \citet{Rodenburg1992}, \citet{Rodenburg1993} and \citet{PENNYCOOK2015160}, both phase and amplitude information can be extracted from a set of diffraction patterns (usually restricting to the Ronchigram region) by performing the Fourier transform of the four-dimensional data set with respect to the scan raster and reordering the dimensions. Depending on the model presumed for the interaction between specimen and incident STEM probe, the direct inversion of the data can either be done by Wigner Distribution Deconvolution (WDD) or the SSB ptychography scheme \citep{Rodenburg1993}. Whereas WDD is based on a single interaction with an arbitrary complex object transmission function and is capable of separating specimen and probe, the weak phase approximation
\begin{equation}\label{eq_wpoa}
    \psi_\text{exit}(\vec{r})=\psi_\text{probe}(\vec{r})\cdot\text{e}^{i\Phi}\approx \psi_\text{probe}(\vec{r})\cdot\left(1+i\phi(\vec{r})\right)\quad.
\end{equation}
governs the SSB approach. Here, the specimen exit wave at a given scan point can be expressed by a multiplication of the probe wave function $\psi_\text{probe}(\vec{r})$ with the first-order Taylor expansion of a phase object with phase distribution $\Phi(\vec{r})$. It is hence limited to weakly scattering ultra-thin objects e.g. thin light matter investigated with relativistic electrons. Any quantitative interpretation of SSB reconstructions of real data should, therefore, be examined critically since eq.~(\ref{eq_wpoa}) breaks down quickly with increasing thickness. Its capability of direct, dose-efficient phase recovery makes it nevertheless attractive for the \textit{in situ} qualitative assessment of specimen and imaging conditions. A more advanced ptychography scheme can still be applied after recording. Because a data point in the final reconstruction depends on all recorded scan points, the reconstruction is only accurate if it is applied to the full 4D data.
\begin{figure}
    \centering
    \includegraphics[width=0.9\linewidth]{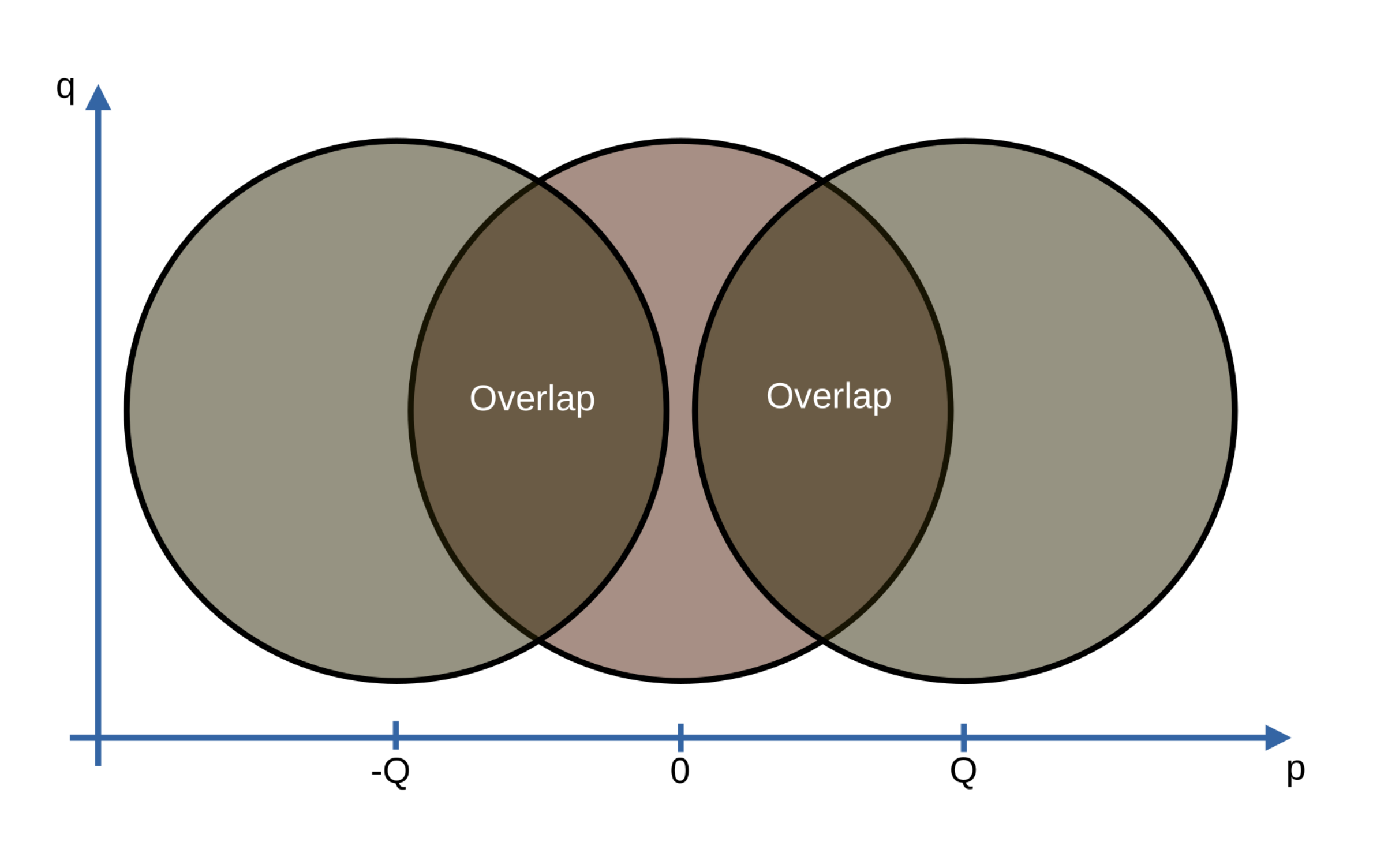}
    \caption{Double overlap regions in the planes of the Fourier transformed 4D data cube are defined by three circles of the size of the probe-forming aperture. The distance $Q$ between the circles is determined by the currently considered spatial frequency of the scan raster and defines the spatial frequency of $\Phi$ reconstructed in this plane of the data cube. Potential triple overlaps for small $Q$ need to be excluded.}
    \label{fig:trotters}
\end{figure}

We refer to original work for a derivation of the conventional SSB methodology~\citep{Rodenburg1992,PENNYCOOK2015160} and give a concise summary here. The first processing step consists of Fourier transforming the 4D data cube as to the scan coordinate, translating the scan coordinate in real space to spatial frequencies $\vec{Q}$ sampled by the scanning probe. This new 4D data cube can be ordered such that each scan spatial frequency defines one plane.  By employing the weak phase object approximation in eq.~(\ref{eq_wpoa}), Rodenburg et al. have shown analytically that the data in each plane is described by three discs of the size of the probe-forming aperture, positioned at the origin and as Friedel pairs at the positions defined by the spatial frequency vector~$\vec{Q}$. Importantly, double overlaps as depicted schematically in Fig.~\ref{fig:trotters} contain the complex Fourier coefficients of $i\Phi$, potentially affected by aberrations of the probe-forming system. The double overlap regions are often referred to as \textit{trotters} colloquially.

The ptychographic SSB reconstruction is a linear function of the input data since the result is obtained with a sequence of linear transformations, such as Fourier transforms, element-wise multiplication, and summation \citep{PENNYCOOK2015160}. Such linear functions are particularly suitable for incremental processing since they are additive. Mathematically, the complete input data can be understood as the sum
of smaller individual data portions that are padded with zeros to fill the shape of the
complete data set. Additive functions allow to calculate the complete result by accumulating the processing results of zero-padded portions in any subdivision
and order. Furthermore, intermediate results can be extracted at any desired stage from the yet incomplete sum of results.

However, directly processing zero-padded data this way is very inefficient for computationally demanding algorithms such as ptychography employing large 4D-STEM data, because the processing effort and memory consumption is amplified by the number of subdivisions.
For that reason, the algorithm should be reformulated to process smaller input data portions without zero-padding.
The additivity and homogeneity of linear functions gives ample freedom to restructure the
underlying data processing flow towards this goal, allowing development of mathematically equivalent implementations that are
optimized for live imaging.

In the particular case of SSB ptychography, individual spatial frequencies of the result $\Phi(\vec{r})$ are extracted from spatial frequencies of signals at specific scattering angle ranges within the double overlaps~\citep{Rodenburg1993,PENNYCOOK2015160} introduced in Fig.~\ref{fig:trotters}.

Suppose we have the intensity of diffraction patterns present as four-dimensional data, which is written as
$\mathbf{D}_{xy} \in \mathbb{R}^{m \times n} \quad \text{where} \quad x \in [s_x], y \in [s_y]$ and the notation $[s_x]$ represents the set of natural numbers not exceeding $s_x-1$. That is, $[s_x] := \{0,1,\hdots,s_x -1\}$ with the total number of elements $s_x$.
The indices $x,y$ represent the scan position index, with the total number of scanning steps being $s_x$ and $s_y$ in each direction. Each matrix $\mathbf{D}$ is a diffraction pattern $\left(d_{pq}\right)$ where $p$ and $q$ represent the pixel index on the detector with dimension $m \times n$. Each pixel index $p q$ corresponds to a scattering angle, or equivalently, spatial frequency in the specimen.

We can write the Fourier transform with respect to the scan raster as
\begin{equation}
\left( f_{pq} \right)_{kl} = \sum_{x = 0}^{s_x-1} \sum_{y = 0}^{s_y-1} \left( d_{pq} \right)_{xy} e^{-i2\pi\left(\frac{xk}{s_x} + \frac{yl}{s_y} \right)}\quad,
\label{eq2}
\end{equation}
where $k$ and $l$ denote the spatial frequencies in the scan dimension, taking the places of $x$ and $y$. Therefore we obtain a four-dimensional dataset in the spatial frequency domain, i.e., $\mathbf{F}_{kl} \in \mathbb{C}^{m\times n}$.

The next step in SSB is to apply a filter $\mathbf{B}_{kl} \in \mathbb{C}^{m \times n}$ for each tuple of spatial frequencies $kl$ that calculates the weighted average of the spatial frequency signal over specific ranges of pixels, i.e. scattering angles $pq$, that can have positive or negative weight as defined by the trotters.
It should be noted that the structure of this filter is different for each tuple of spatial frequencies.

Using \eqref{eq2}, we can write the reconstruction in the Fourier domain $p_{kl}$ as
\begin{equation}\label{eq_mkl}
\begin{aligned}
p_{kl} &= \sum_{p = 0}^{m-1} \sum_{q = 0}^{n-1} \left(f_{pq} \right)_{kl} \left( b_{pq} \right)_{kl}\\
&=
    \sum_{p = 0}^{m-1} \sum_{q = 0}^{n-1} \left[\sum_{x = 0}^{s_x-1} \sum_{y = 0}^{s_y-1} \left( d_{pq} \right)_{xy} e^{-i2\pi\left(\frac{xl}{s_x} + \frac{yk}{s_y} \right)}\right]\left( b_{pq} \right)_{kl}\\
    &= \sum_{x = 0}^{s_x-1} \sum_{y = 0}^{s_y-1} \left[\sum_{p = 0}^{m-1} \sum_{q = 0}^{n-1}\left( b_{pq} \right)_{kl}\left( d_{pq} \right)_{xy} \right] e^{-i2\pi\left(\frac{xk}{s_x} + \frac{yl}{s_y} \right)}\;.
    \end{aligned}
\end{equation}
The reformulation is given by interchanging the summation and the index dimension of detector and the index of scan points in the real and frequency domain, respectively. This nested summation can be pruned without changing the result by skipping parts that are known to yield zero, for example calculations for empty double overlap regions. Note that $p_{kl}$ essentially represents the Fourier coefficients of $i\Phi_{xy}$ in eq.~(\ref{eq_wpoa}) which are potentially affected by aberrations of the probe-forming system. 

The inner part of  eq.~(\ref{eq_mkl}) can be implemented with a matrix product between $\mathbf{B}$ and $\mathbf{D}$. The outer sum over $x$ and $y$ can then be sub-divided and reordered to process the input data $\mathbf{D}$ incrementally in smaller portions. Numerically, the filter matrix $\mathbf{B}_{kl}$ can be stored efficiently as a sparse matrix since the trotters for the highest frequencies are often empty (no overlap), and for many frequencies the double overlap region where the filter is non-zero, is small.

As an intermediate summary, this formulation translates the SSB scheme to a matrix product of a partial input data matrix
with a sparse matrix containing the double overlap regions. Elements of a Fourier transform are then applied to the result of this matrix product. This generates a partial reconstruction result in the frequency domain that covers the entire field of view.
The partial reconstructions for all partial input data portions are accumulated in a global buffer using a sum, as described above.
For a live view of the reconstruction in the spatial domain, the contents of this buffer
can be inversely Fourier transformed at any desired time to show the transmission function of the specimen within the weak phase object approximation.

We used LiberTEM \citep{clausen_alexander_2020_3982290} as a data processing framework since it is optimized for MapReduce-like
approaches and designed with live data processing capabilities in mind \citep{Clausen2020}. \mbox{LiberTEM} user-defined functions (UDFs) provide
the application programming interface (API) to efficiently implement operations that follow the described pattern: A method
to process a stack of frames that is called repeatedly, task data to store constant data such as the sparse
matrix for the double overlaps, result buffers of arbitrary type and shape, and user-defined merging operations to generate a
result of arbitrary complexity from partial results. Furthermore, LiberTEM allows to update a result display each time a partial
result is merged. UDFs for first moment analysis and virtual detectors were already implemented before
for offline data analysis.

To run the UDFs on live data, we implemented a prototype live UDF back-end that allows to run a set of UDFs on
data from a Quantum Detectors Merlin for EM Medipix3 for electron microscopy \citep{PLACKETT20130123,QUANTUMDETECTORS201911}. It uses multiple CPU cores for decoding the raw data from the detector, and a single GPU for the main processing task, which was sufficient
for this application. A production-ready version with support for multiple processing nodes, further enhanced use of multiple CPUs
and multiple GPUs similar to the offline data processing capabilities of LiberTEM is being designed at the time of writing and will be published as open-source as a part of LiberTEM.

For comparison, ePIE \citep{MAIDEN200909} based ptychographic reconstructions have been performed in post processing as it can reconstruct both the complex object transmission function and the complex illumination, still presuming single interaction of probe and specimen, but considering an arbitrary complex object transmission function. This was done to check whether the straightforward SSB approach compromises the quality of the result compared to more advanced ptychographic schemes. It has to be noted that, as an iterative method, ePIE is not suitable for live imaging in the current realisations.

\subsection{Material system}
We demonstrate live processing using 
two different specimens.
First, 
indium selenide (In$_2$Se$_3$) was used to highlight the advantages of live ptychography and
centre of mass in comparison to conventional STEM \citep{YE1998}.
Second, a strontium titanate (SrTiO$_3$) lamella with the electron beam incident along the [100] axis for a more quantitative analysis was used. The latter is stable and provides good contrast in conventional STEM for comparison and adjustments, is well-characterized, and at the same time can highlight the ability to also resolve the light oxygen columns that are difficult to image with reliable contrast by conventional STEM techniques \citep{BROWNING1995}.
Too small double overlap regions with an area of less than 10\,px were omitted in live processing to avoid the introduction of excessive noise. This can be understood as a band-pass filter applied to exclude the high frequencies close to the transfer limit of SSB ptychography which corresponds to twice the radius of the probe-forming aperture.

\subsection{Experimental setup}
First moment imaging and ptychography require the correct alignment and calibration of the scan (specimen) coordinate system with respect to the detector coordinate system. The convergence angle of the incident probe was measured with a polycrystalline gold specimen. Employing parallel illumination first, the (111) gold diffraction ring was used to calibrate the diffraction space assuming a lattice constant of gold of 0.4083\,nm \citep{Villars2016:sm_isp_sd_1628821}. With the known wavelength the convergence semi-angle was determined to 22.1\,mrad from a Ronchigram recorded in the same STEM setting as used in the actual experiment.

The rotation of the detector coordinate system with respect to the scan axes was determined by minimizing the curl of the first moment vector field and making sure that the divergence of the field is negative at atom positions. Note that, in theory, the curl of purely electrostatic fields should vanish. The pixel size in the scan dimension was taken from the STEM control software during live processing and verified by comparison with the known lattice constant of SrTiO$_3$. The residual scan distortion, that is, the translation of the diffraction pattern as a whole during scanning, was not compensated for since it turned out to be negligible at the atomic-resolution STEM magnifications used in this analysis.

Data was acquired at a probe-corrected FEI Titan 80-300 STEM \citep{HEGGEN20160201} operated at $300\,$kV. The microscope was equipped with a Medipix Merlin for EM detector operated at an acquisition rate for a single diffraction pattern of $1$\,kHz in continuous mode. The scan size was $128\times128$ scan points and the recorded diffraction patterns had a dimension of $256\times256$ pixel. In addition, the high-angle annular dark field (HAADF) signal has been recorded with a Fischione Model~3000 detector covering an angular range of 121.7\,mrad to approximately 200\,mrad. The upper limit is defined by apertures of the microscope rather than by the outer radius of the detector. The run time for each step of the data processing flow was measured using the {\tt line-profiler} Python package.

\section{Results}
\begin{figure}
    \centering
    \includegraphics[width=\linewidth]{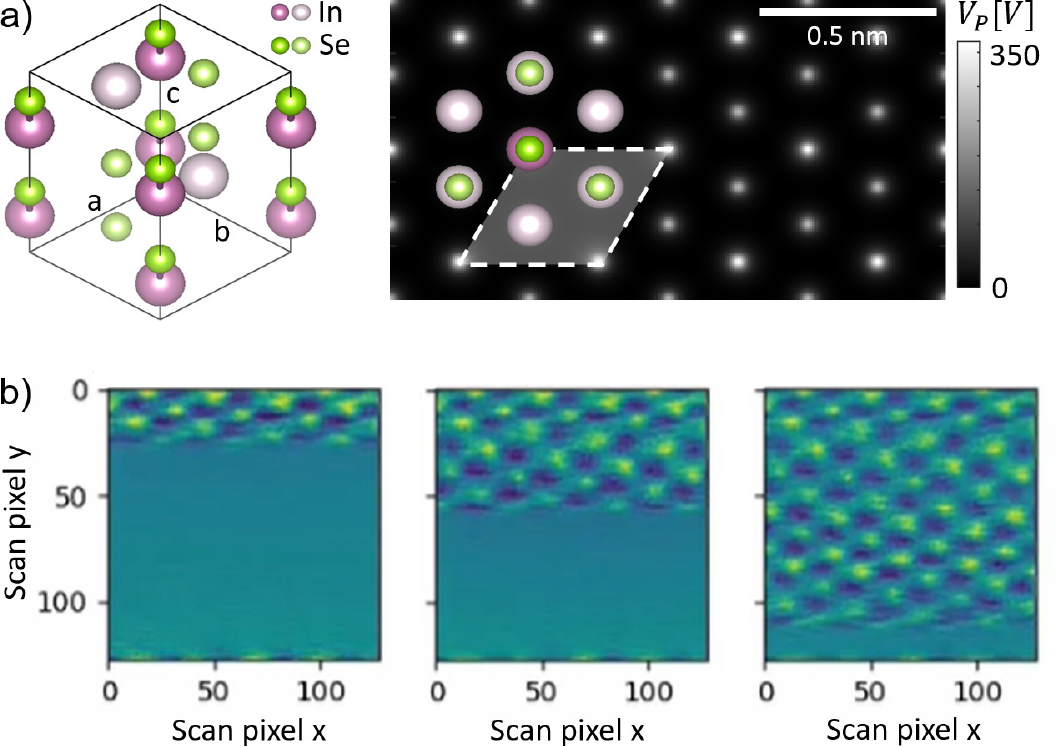}
    \caption{Investigation of In$_2$Se$_3$ nanosheets. (a)~3D view of the unit cell (left) and electrostatic potential projected along $\vec{c}$. Note the different densities of atoms in c-direction at different sites. (b)~Snapshots of live ptychography (phase) using the SSB scheme at different stages of the scan.}
    \label{fig:livesteps}
\end{figure}
\subsection{Live processing}
The performance of our approach is demonstrated in Fig.~\ref{fig:livesteps} which depicts (a)~the structure of In$_2$Se$_3$ together with the projected potential and (b)~snapshots of the reconstructed phase in nanosheets of this material using SSB ptychography at different stages of the scan. The reconstruction is performed progressively such that only the diffraction patterns of incoming scan pixels are processed and added to the final reconstruction according to~eq.\,\eqref{eq_mkl} in which all pixels are affected. Whereas current research in the field of materials science explores ultra-thin In$_2$Se$_3$ as a candidate for 2D ferroelectrics, this specimen was chosen due to its robustness against electron dose for the methodological development. The data in Fig.~\ref{fig:livesteps}\,b was taken from a live video recorded during the STEM session, being available as supplementary material online.

\begin{figure}
 \centering
 \includegraphics[width=\linewidth]{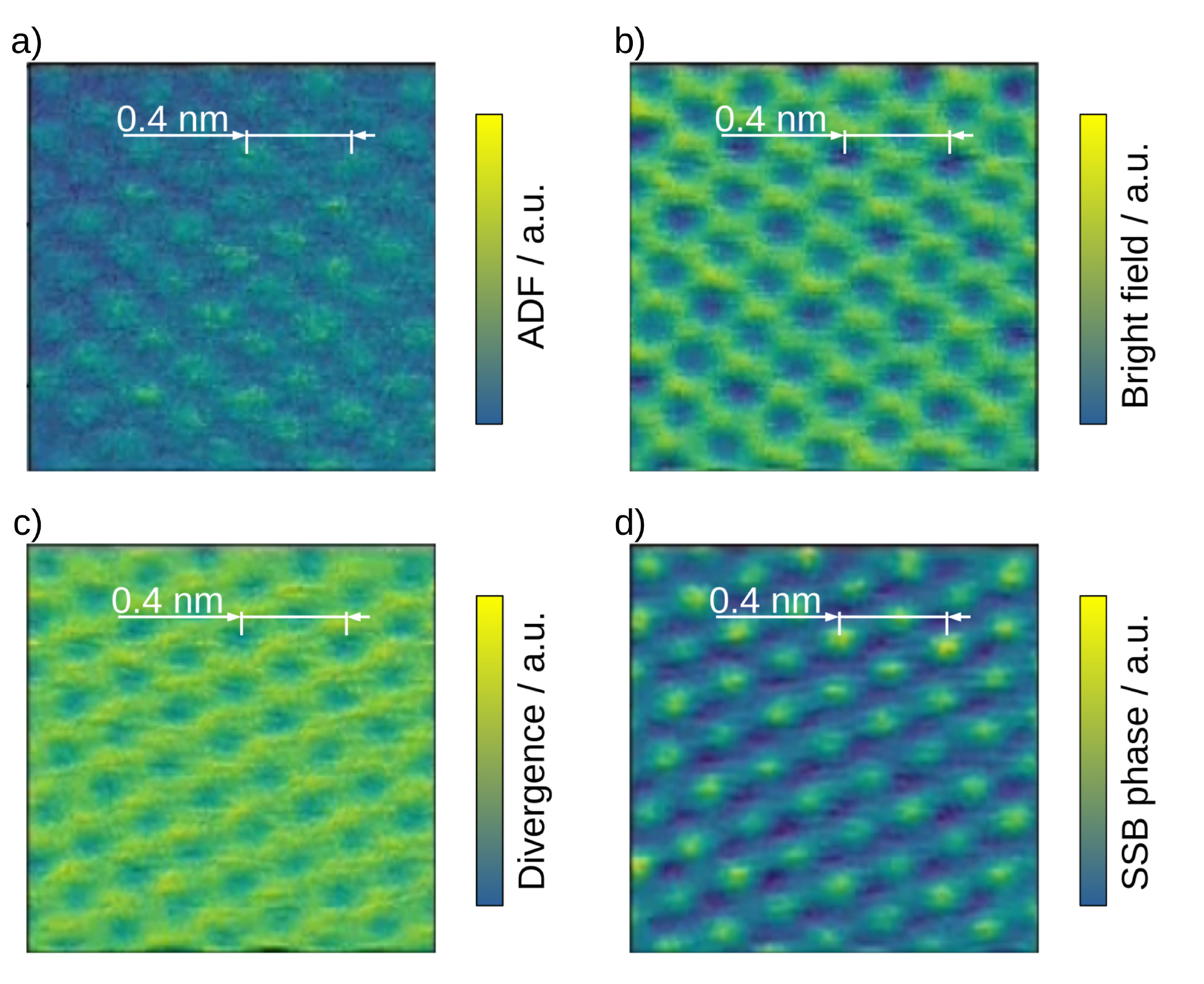}
 \caption{Live imaging of In$_2$Se$_3$ nanosheets at atomic resolution for a $128\times128$ STEM scan. The atom columns are visible in all methods. The semi-convergence angle was 22.1\,mrad.}
 \label{fig:In2Se3}
\end{figure}
In general, the quality of the ptychographic reconstruction strongly depends on matching the position and radius of
the (usually circular) aperture function on the detector with the double-overlap regions. Even small deviations lead to a mismatch between the mask borders and the edges of the zero-order disk, which adds noise and reconstruction errors from those erroneous border pixels. A precise alignment can be done, e.g. using a position-averaged diffraction pattern or by recording the bright-field disc with the specimen removed from the field of view, as in our analyses.

It must be pointed out that a partial SSB reconstruction only approximates the result because not all spatial frequencies have been completely sampled at this point. However, it is already possible to see the atom columns in the partial reconstructions. This allows time- and dose-efficient live focusing and navigation to regions of interest on specimens. In Fig.~\ref{fig:In2Se3}, complete reconstructions of In$_2$Se$_3$ are shown. All atom columns can be seen clearly in the bright field images, in the divergence of the first moment and in the phase of the object transmission function obtained by SSB ptychography. All signals were calculated live from the 4D-STEM data stream.

\begin{figure}
    \centering
    \includegraphics[width=\linewidth]{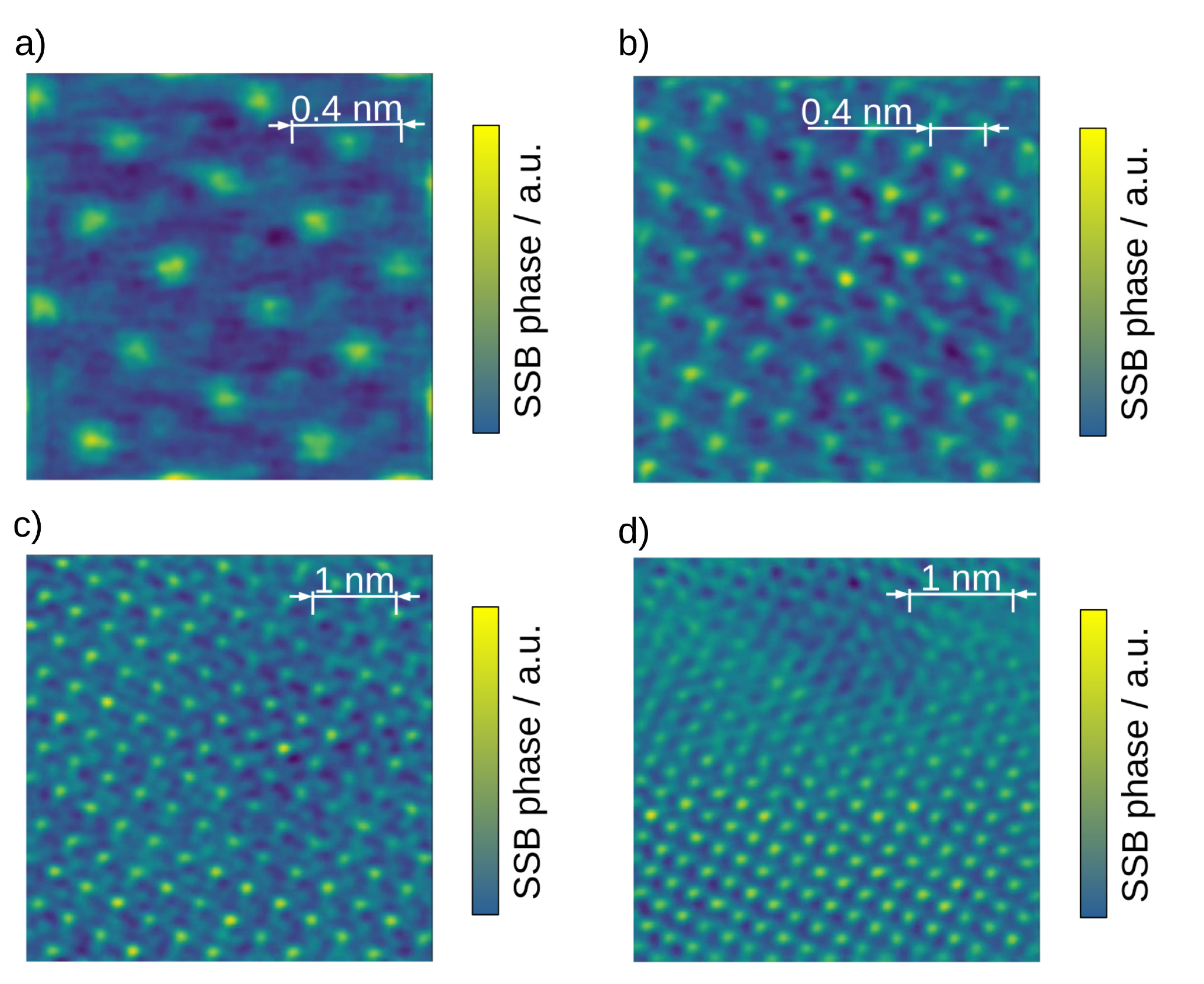}
    \caption{Live variation of microscope parameters on SrTiO$_3$: During live reconstruction the parameters for SSB were not changed. From (a) to (c) the magnification is reduced. If the magnification is reduced further, SSB fails. In (d) the focus is changed during acquisition. The semi-convergence angle was 22.1\,mrad.}
    \label{fig:liveVariation}
\end{figure}
Figure~\ref{fig:liveVariation} depicts the phase of ptychographic SSB reconstructions of SrTiO$_3$ performed live while changing microscope parameters such as the STEM magnification (scan pixel size) in Fig.~\ref{fig:liveVariation}\,a-c, and the probe focus in Fig.~\ref{fig:liveVariation}\,d. The full video showing live updating during continuous scanning while microscope parameters are changed is available in the supplementary material online.

Although the inherent reconstruction parameters are naturally robust against a change of the probe focus for direct ptychography schemes that do not intend to correct aberrations, a magnification change influences the relation between a spatial frequency in the reconstruction in pixel coordinates, and the pixel distance in sample coordinates. This means that the geometry of a double overlap region that contains the signal to reconstruct a certain spatial frequency in the result changes with magnification. In the strict sense, changing the magnification without updating the scan step size (and double overlap masks) in the reconstruction is, therefore, mathematically inaccurate. In practice, however, it is desirable to be able to lower the magnification so as to navigate coarsely to a specimen region of interest without a new calibration or pre-calculation of the masks $\mathbf{B}_{kl}$ in eq.~(\ref{eq_mkl}) to be time-efficient, and then observe or record details after switching back to the correct magnification again. Therefore, we studied to which extent the reconstruction is robust against a magnification change despite keeping the internal SSB parameters unchanged in Fig.~\ref{fig:liveVariation}\,a-c. It showcases that the structural contrast is in general maintained. A possible future development with only a moderate effort could be pre-defining double overlap masks for common magnifications in order to build a look-up table that solves inaccuracies. We assume without loss of generality regarding the algorithmic structure, that live processing is required at a dedicated magnification being equal to the final recording where data is actually stored, and mention a certain robustness against changes of the scan pixel size as a valuable side note.

\subsection{Computational details}
In our implementation the sparse matrix product was the throughput-limiting step. By using an efficient GPU implementation from cupyx.scipy.sparse (\citeauthor{Cupy}), this could be accelerated sufficiently to allow processing of over 1000 frames per second at suitable parameters, enabling for live imaging using a Medipix3 sensor.

The memory consumption of the current SSB ptychography implementation scales as $\mathcal{O}\left(N\right)$ with the number of scan points at constant aspect ratio since the number of spatial frequencies to reconstruct scales linearly with the number of scan points, and each reconstructed frequency adds a double overlap region. These regions sample the diffraction data more densely when extracting more spatial frequencies, meaning the average number of non-zero entries per trotter is roughly constant for a given pixel size and beam parameters. As the processing time for dense-sparse matrix products roughly scales linearly with the number of non-zero entries in the sparse matrix and number of vectors in the dense matrix, SSB scales poorly with $\mathcal{O}\left(N^2\right)$ in computation time with the number of scan points.

The size of the matrix containing the double overlap regions is highly dependent upon the acquisition parameters in the present implementation. A larger camera length increases the Ronchigram size, consequently increasing non-zero entries.
The relationship between scan pixel size and the convergence angle changes the spatial frequency limit above which no double overlaps occur. The less often spatial frequencies create double overlaps, the smaller the matrix $\mathbf{B}$ becomes. In this study the scan area for live processing was limited to a size of $128\times128$ and microscope parameters were chosen to keep the matrix size low enough to fit into the GPU RAM.

\subsection{Post processing}
We concentrated up to this point, solely on the live evaluation in order to facilitate optimisation of experimental parameters during the session. This avoided the need for saving vast amounts of 4D-STEM data that would have required significant disc space due to the continuous nature of the experiments. During the experiments only a few reliable data sets were recorded.
To verify the reliability of our approach we compared the post-processing of the recorded data against the live experiment to determine if the inherent parameters used in post-processing such as the Ronchigram position and radius, are in sufficient agreement with those of the live results.

\begin{figure}
    \centering
    \includegraphics[width=\linewidth]{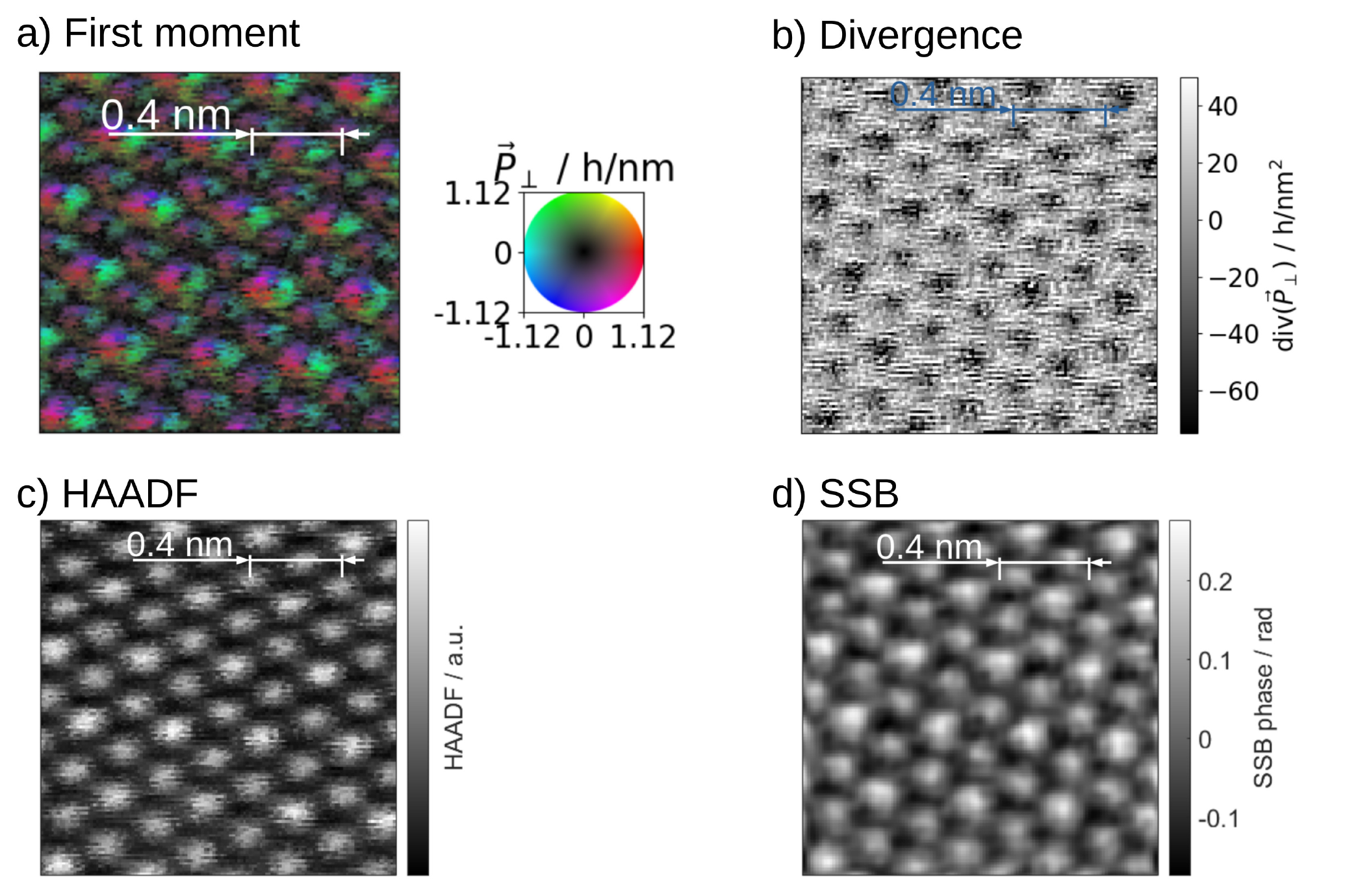}
    \caption{Post processing of In$_2$Se$_3$. In all images, the positions of the two different atom columns of In$_2$Se$_3$ can be seen.}
    \label{fig:In2Se3post}
\end{figure}

In Fig.~\ref{fig:In2Se3post}, the post processing of In$_2$Se$_3$ data is shown. Here, the two different atom columns highlighted in Fig.~\ref{fig:livesteps}\,a can be seen in the first moment vector field (Fig.~\ref{fig:In2Se3post}\,a), in its divergence (Fig.~\ref{fig:In2Se3post}\,b), and in the HAADF image (Fig.~\ref{fig:In2Se3post}\,c). The phase of the SSB reconstruction (Fig.~\ref{fig:In2Se3post}\,d) also yields site-specific contrast, which is more pronounced than in the live imaging result in Fig.~\ref{fig:In2Se3}\,d. Recalling that SSB ptychography relies on the weak phase object approximation in eq.~(\ref{eq_wpoa}) that breaks down already at the thinnest of specimen as to a quantitative interpretability, conclusions from relative phases among different atomic sites need to be drawn with great care.

\begin{figure}
    \centering
    \includegraphics[width=\linewidth]{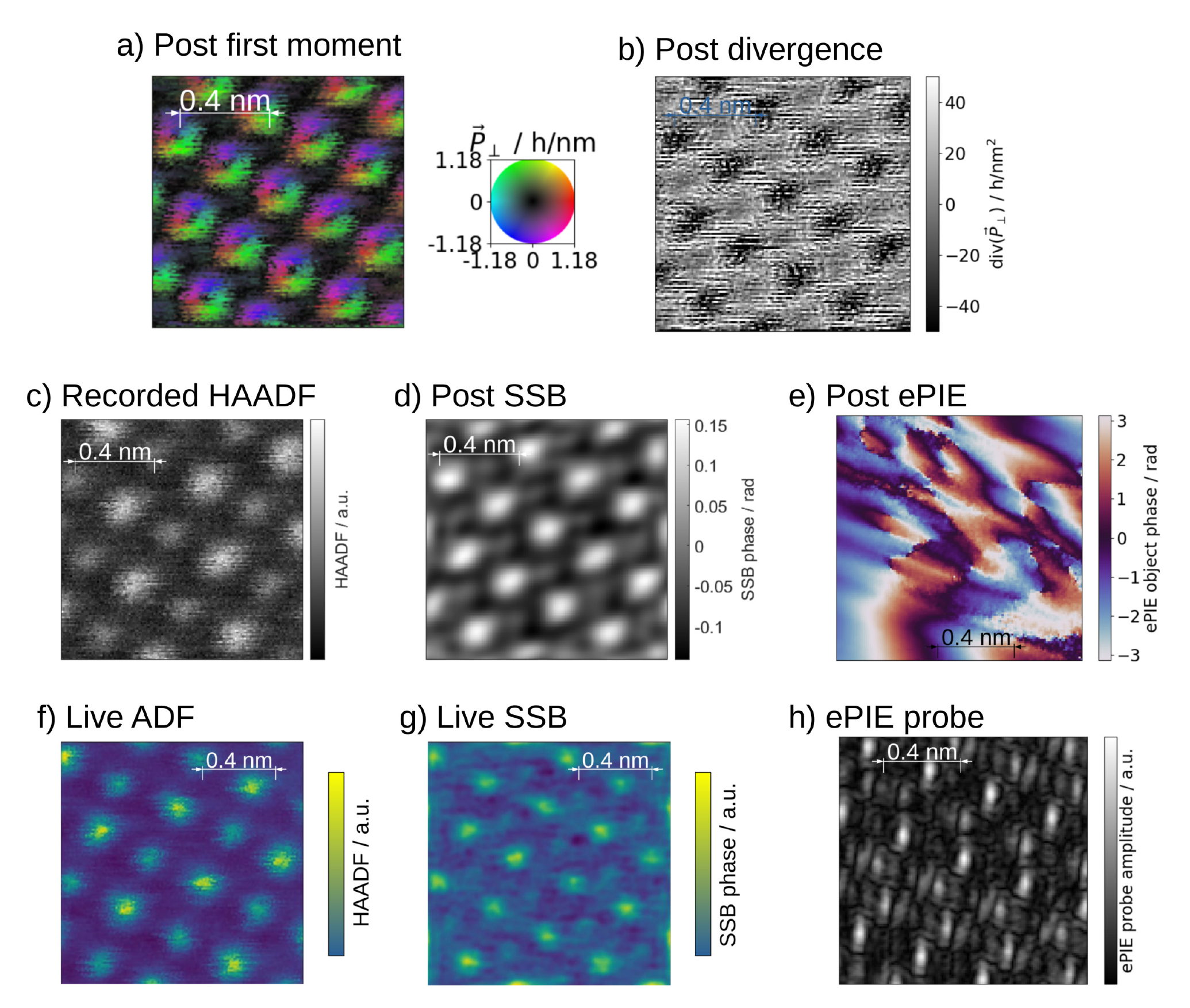}
    \caption{Comparison between live imaging and post processing on SrTiO$_3$. In all images except~(e), the positions of the strontium and the titanium oxide atom columns are clearly visible. The position of the oxygen columns can only be seen in the SSB results~(d, g) and in the divergence of the first moment~(b). The ePIE result does not show any information in the phase object. (h) The amplitude of the probe from the ePIE reconstruction shows some lattice information. The ePIE reconstruction has a 4° rotation compared to the other post processed results due to the rotation angle. This owes to the implementation of dealing with the scan rotation. The semi-convergence angle was 22.1\,mrad, the sample thickness was approximately 25\,nm, determined by comparing the PACBED with simulation as shown in Figure~\ref{fig:PACBED}.} 
    \label{fig:STOliveVsPost}
\end{figure}
In Figure~\ref{fig:STOliveVsPost}, a comparison between different signals as well as between live imaging and post processing of SrTiO$_3$ is shown. Strontium titanate enables evaluation of the different imaging modes regarding their capability for  simultaneous imaging of light oxygen columns and comparably heavy Sr and Ti oxide columns, adding to the results obtained for In$_2$Se$_3$ in Fig.~\ref{fig:In2Se3post}. The first moment vector field (Fig.~\ref{fig:STOliveVsPost}\,a) predominantly shows the heavy atom columns as sinks. The fact this vector field also contains sinks at the oxygen sites becomes visible in the divergence map in Fig.~\ref{fig:STOliveVsPost}\,b, whereas the HAADF signal recorded separately with the conventional annular detector in Fig.~\ref{fig:STOliveVsPost}\,c visualises only the heavy-atom sites of Sr and Ti oxide. In the post-processed SSB~(Fig.~\ref{fig:STOliveVsPost}d) the oxygen columns can be easily determined simultaneously with the heavy sites. The live  SSB~(Fig.~\ref{fig:STOliveVsPost}g) has some noise, but the oxygen columns are still visible. Moreover, a comparison of the conventional HAADF in Fig.~\ref{fig:STOliveVsPost}\,c with the annular dark field signal generated by a virtual annular detector applied to the 4D-STEM data in Fig.~\ref{fig:STOliveVsPost}\,f demonstrates that practically all main contrast mechanisms exploiting low- as well as high-angle scattering can be captured by the 4D-STEM imaging mode. 
\begin{figure}
    \centering
    \includegraphics[width=\linewidth]{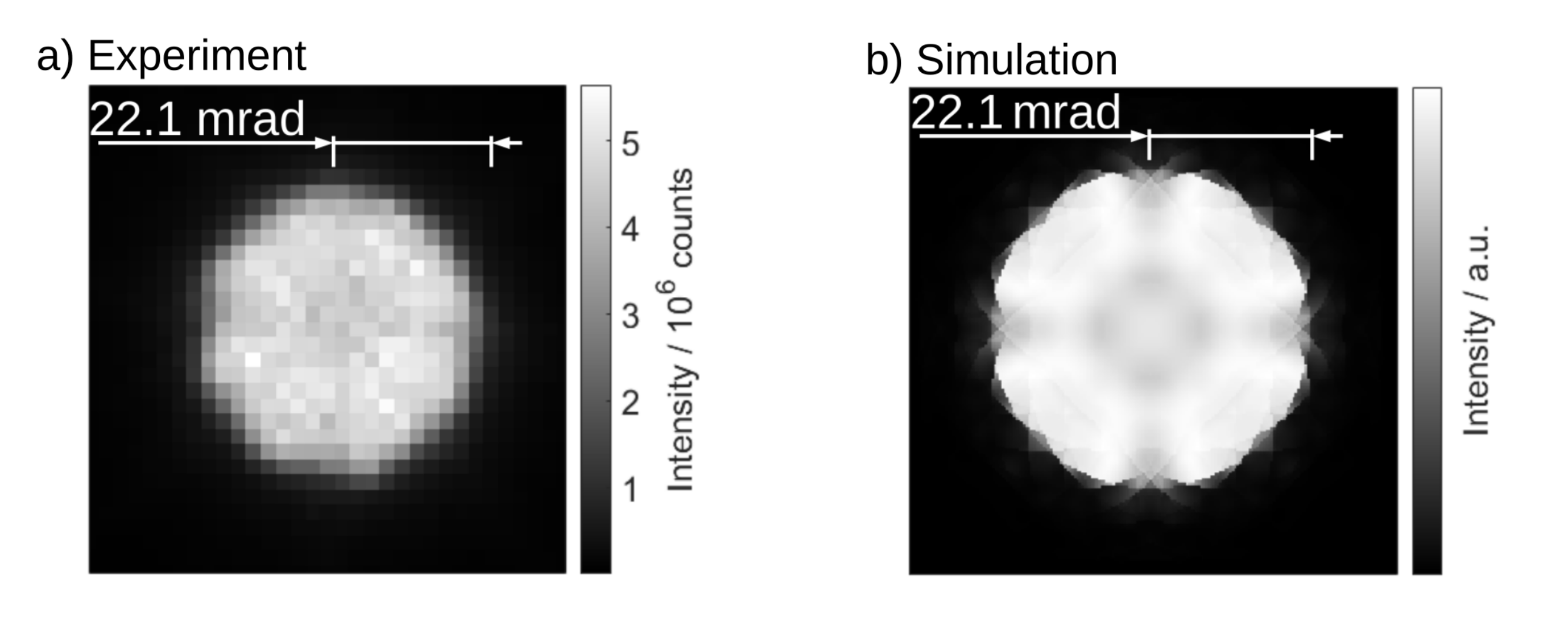}
    \caption{Comparison of the PACBED of the scan in Figure~\ref{fig:STOliveVsPost} with simulation: The best match is achieved at a simulated thickness of 25\,nm.}
    \label{fig:PACBED}
\end{figure}

Earlier we stated that the live ptychographic reconstruction exploits the SSB algorithm, because it is a non-iterative, linear and direct scheme that allows for in-situ processing. 
However, the weak phase object interaction model given by eq.~(\ref{eq_wpoa}) is a seemingly drastic limitation of this approach.
In the analysis of post processing data, it is beneficial to explore whether a ptychographic scheme based more on an advanced interaction model, neglecting computational hardware constraints for the moment, could have been the better choice for live ptychography. To this end, we used the ePIE algorithm to reconstruct the SrTiO$_3$ data as shown in Fig.~\ref{fig:STOliveVsPost}\,e. The standard ePIE implementation clearly does not give a reasonable result, at least for the usual reconstruction settings reported in literature that we used here. To explore this in more detail, the sample thickness was determined to approximately 25\,nm by comparing the experimental PACBED with a thickness dependent simulation as in Fig.~\ref{fig:PACBED}. Consequently, the specimen thickness was far beyond the validity of both the weak phase object approximation and the single-slice model used in ePIE. At first sight it is nevertheless surprising that the latter approach performs worse, since one must consider it more advanced than the weak phase model from the viewpoint of scattering theory. In fact, ePIE tries to iteratively find both the probe and the object transmission function in such a way that the modulus of the Fourier transform of their product agrees best with all details of the experimental diffraction data. When dynamical scattering sets in, this multiplicative interaction scheme is incapable of delivering the details of the diffraction pattern, so that the algorithm does not converge to a reliable solution. In this particular case ePIE has put some lattice information in the probe (Fig.~\ref{fig:STOliveVsPost}\,h). This will be studied in the next subsection by means of simulations.

To summarise the experimental results, performing live ptychography using the SSB method had originally been motivated by computational aspects, but contrary to expectations it also turns out to be more robust against the violation of the weak phase object approximation and dynamical scattering. Of course, this only holds for qualitative imaging, but it is a significant advantage in practice where suitable structural contrast is obtained also at elevated specimen thickness for both light and heavy atomic columns.

\subsection{Simulation studies}
\textbf{Partial reconstruction.~~}
\begin{figure}
    \centering
    \includegraphics[width=\linewidth]{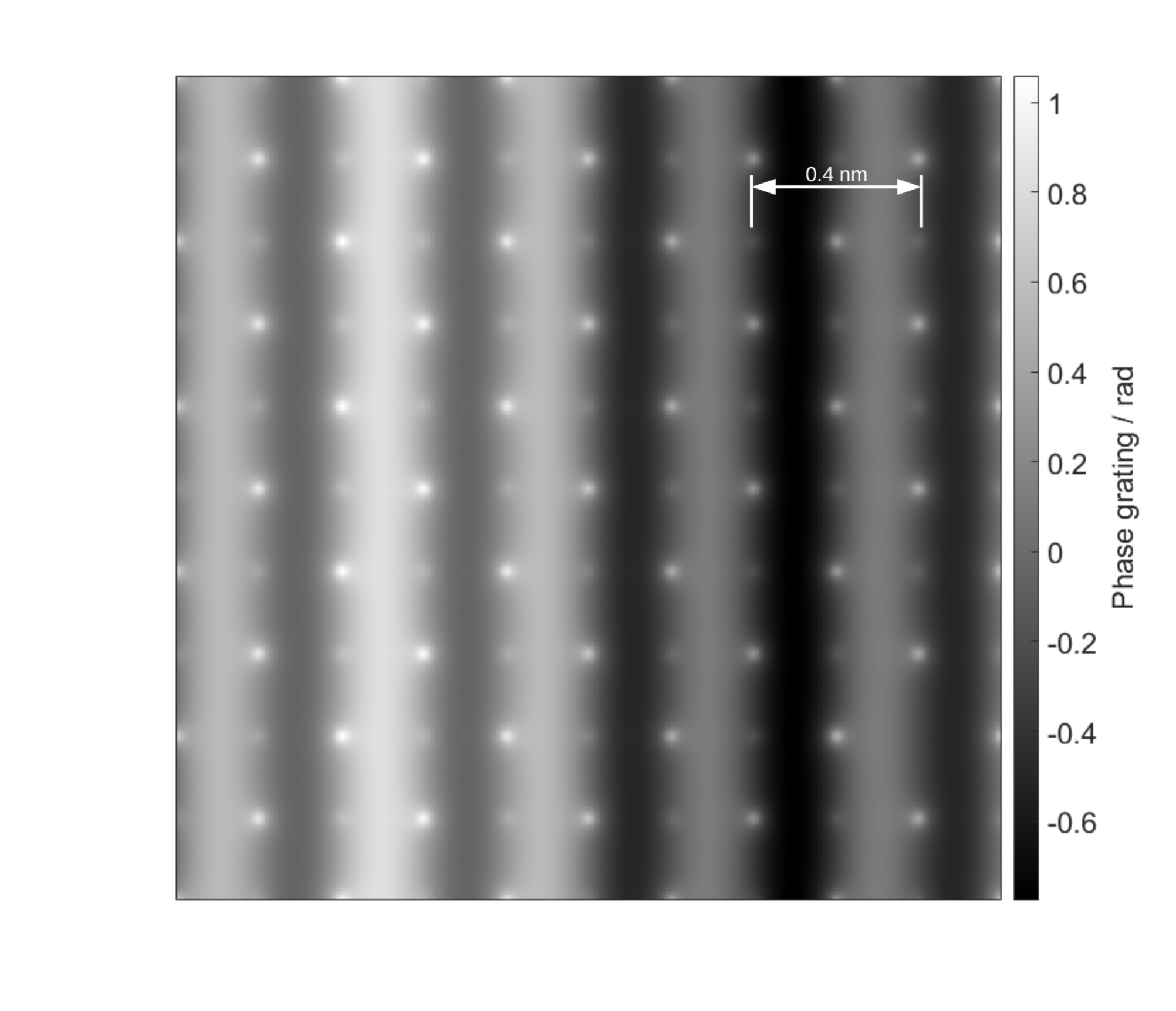}
    \caption{Phase grating that was used for the simulation for the partial SSB-reconstructions in Figure~\ref{fig:progressive}.}
    \label{fig:progressivePG}
\end{figure}
\begin{figure}
    \centering
    \includegraphics[width=\linewidth]{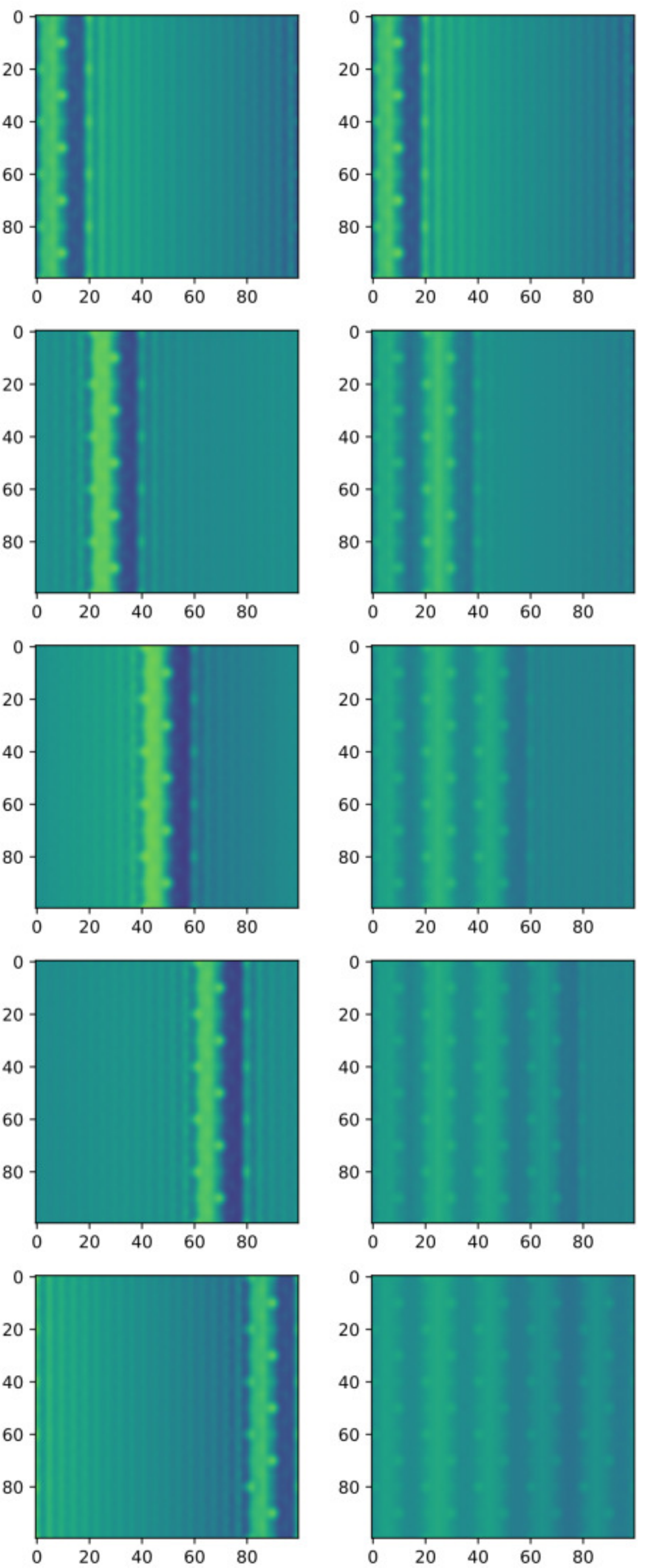}
    \caption{Visualization of accumulation of partial reconstructions using a synthetic dataset of simulated SrTiO$_3$ combined with long-range potential modulation. The left column shows the reconstruction of disjoint subsets of the input data, and the right one the accumulated result until all data is processed and the complete result is obtained. An animation with smaller subdivisions is available in the supplementary material.} 
    \label{fig:progressive}
\end{figure}
 As the implementation of live ptychography based on eq.~(\ref{eq_mkl}) maps the result of single scan points successively to the final reconstruction, a simulation study has been conducted in which we investigated the accuracy of the reconstructed phase in already scanned regions in dependence of the scan progress. A synthetic dataset has been simulated, based on an SrTiO$_3$ unit cell as a starting point. Then, a five by five super cell was created by repetition and the phase grating (Fig.~\ref{fig:progressivePG}) has been calculated. Two artificial spatial frequencies were added to the phase grating, one with a wavelength of a single unit cell and one with a wavelength of the super cell. To eliminate dynamical scattering in this conceptual study, a 4D-STEM simulation with $20\times20$ scan points per unit cell was performed using only one slice with a thickness of one unit cell along electron beam direction $[001]$. Finally, partial SSB reconstructions have been performed using the full range of scan pixels in vertical directions, but only portions of 20 scan pixels horizontally, mimicking a reconstruction during progressive scanning as depicted in Fig.~\ref{fig:progressive}. An animation calculated from 100 single pixel columns is available in the supplementary material.

Figure~\ref{fig:progressive} contains the individual blocks of the 20 scan pixel wide reconstructions in the left hand column, and the accumulated result on the right hand side with the scan progressing from top to bottom. Consequently, the phase bottom right is the final reconstruction for the full scan, obtained by our cumulative approach which we found to be identical to a reconstruction using the conventional treatment employing the whole 4D scan. Only here, the low, medium and atomic-scale spatial frequencies are reconstructed correctly without artefacts. In that respect, it is instructive to explore the partial reconstructions in Fig.~\ref{fig:progressive}. Since we used subsets that equal the size of a single unit cell, spatial frequencies down to the synthetic one with a period of one unit cell appear at least qualitatively in all partial reconstructions. The sharp edges between the available data and the yet-missing region have resulted in a ringing effect near the edges known as the Gibbs-phenomenon. A closer look at the left column of Fig.~\ref{fig:progressive} exhibits that these artefacts largely interfere destructively during accumulation as seen, e.g. by a maximum at horizontal scan pixel 20 in the top row and a minimum at this position in the row below. Consequently, ringing artefacts become less obvious in the full reconstruction on the right within the region that has already been scanned. Therefore, it is already possible to visualise the atomic structure and partly meso-scale phase variations for partial scans, making it possible to navigate on the sample and visually interpret results in a real experiment.

\begin{figure}
    \centering
    \includegraphics[width=\linewidth]{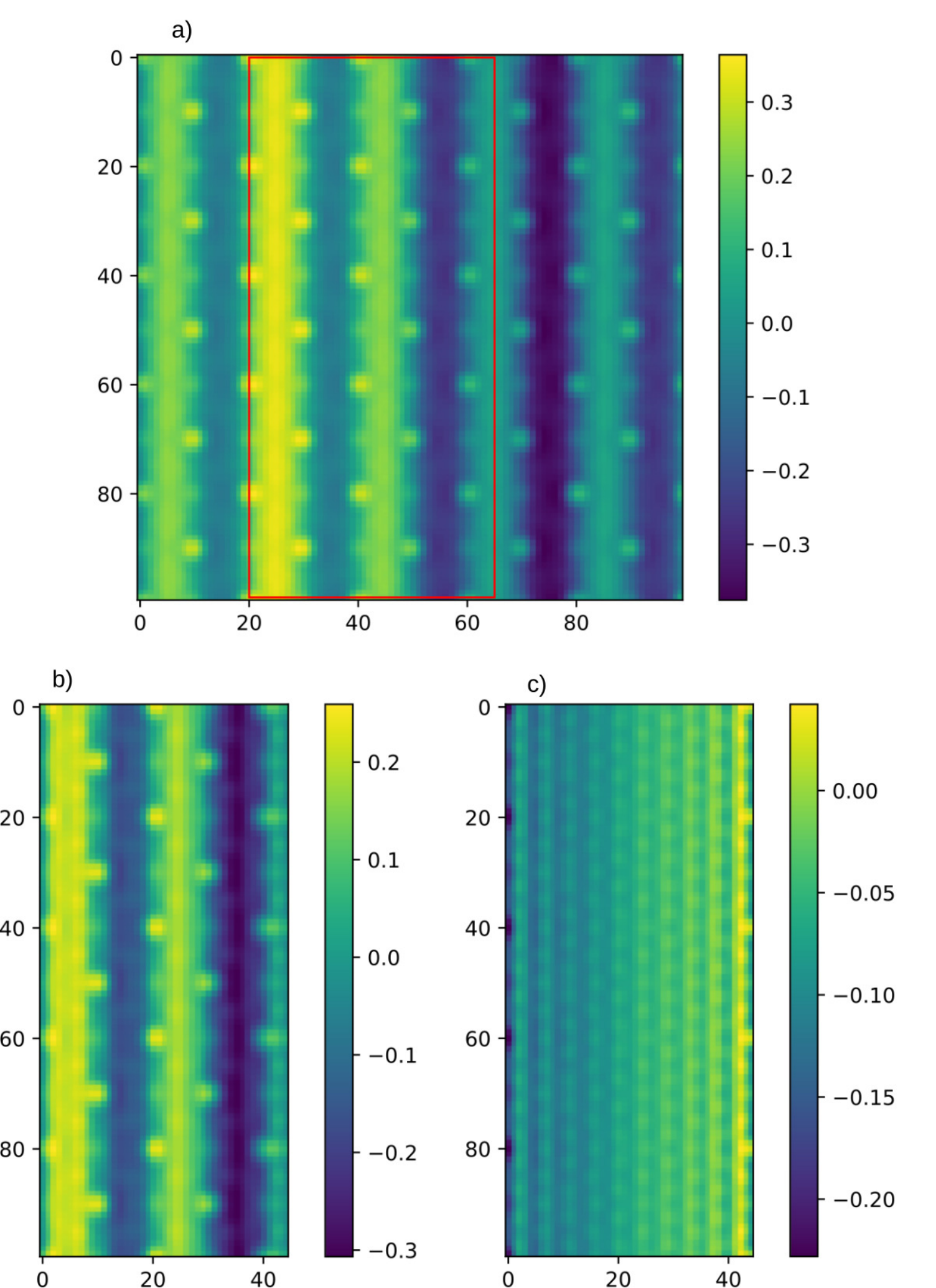}
    \caption{Reconstruction limited to a cutout from the data in both reconstructed area and input data, as opposed to partial reconstruction of the area of the full dataset in Figure~\ref{fig:progressive}. (a) shows the selected cutout area from the full reconstruction, (b) the reconstruction limited to this area, and (c) the difference between the two. This demonstrates how a discontinuity from wrapping around at the edges for a periodic boundary condition creates reconstruction artefacts due to the high frequency cut-off of SSB.}
    \label{fig:cutout}
\end{figure}

Real specimens and scan regions do not usually fulfill periodic boundary conditions which still apply to the full scan of Fig.~\ref{fig:progressive}. Therefore, Fig.~\ref{fig:cutout} shows the impact of selecting different reconstruction areas by simulating the reconstruction of a smaller scan area that is not aligned with the underlying lattice. SSB reconstructs the specimen with an assumption of a periodic boundary condition and cannot reconstruct spatial frequencies above a certain threshold, as previously discussed. Trying to reconstruct a field of view where wrapping around the edges creates a discontinuity, i.e. frequencies higher than SSB can reconstruct, leads to reconstruction artefacts as seen in Fig.~\ref{fig:progressive}. Quantitatively, the difference between the ground truth taken from the marked rectangle of the full reconstruction in Fig.~\ref{fig:cutout}\,a and a reconstruction that solely employs the scan pixels therein, as seen in Fig.~\ref{fig:cutout}\,b is mapped in Fig.~\ref{fig:cutout}\,c and can take significant values of 5-10\,\% of the phase itself in the present example.

As a solution, the field of view where an accurate reconstruction is required can be surrounded by a smooth transition to a zero-valued buffer area so that the presence of spatial frequencies above the resolution limit is minimized when the reconstruction area is wrapped around at the edges. In future studies a Lanczos filtering \citep{Duchon1979} scheme could be added for our live ptychography approach.

\begin{figure}
    \centering
    \includegraphics[width=\linewidth]{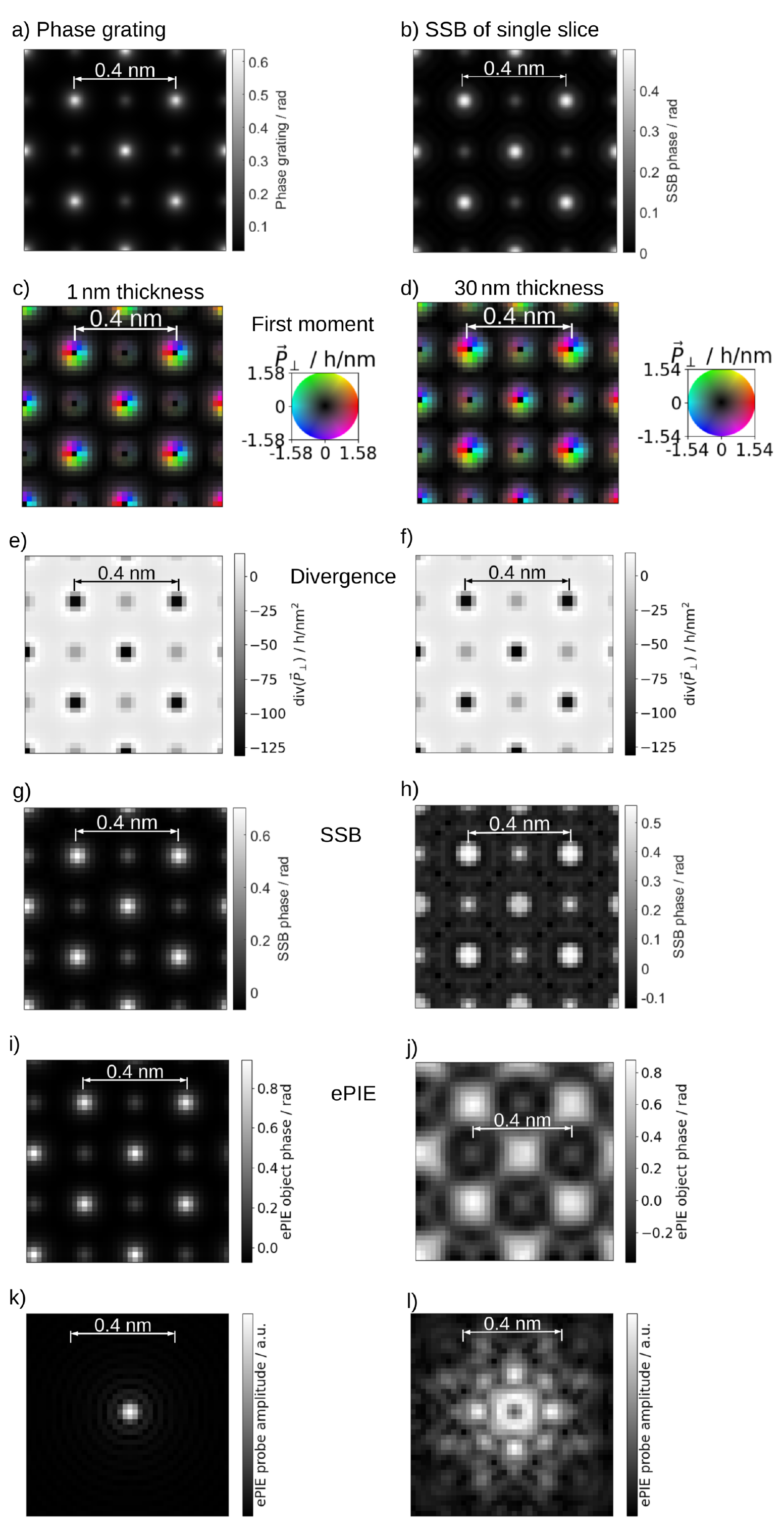}
    \caption{Simulations for SrTiO$_3$: (a) Phase grating used for multislice simulation, (b) SSB from a single slice simulation using also the weak phase approximation in the simulation, (c, d) First moment, (e, f) Divergence of first moment, (g, h) SSB without using weak phase approximation in the simulation of the 4D-STEM-data, (i, j) phase object from ePIE, (k,l) probe amplitude from ePIE, (c, e, g, i, k) 1\,nm sample thickness and (d, f, h, j, l) 30\,nm sample thickness. The simulation parameters where chosen to match those used in the experiments. Additionally the following parameters were used: 22.1\,mrad semi-convergence angle and 20 by 20 scan points per unit cell.}
    \label{fig:SimulationSTO}
\end{figure}

\textbf{Thickness effects.~~}
A comprehensive algorithmic review is not our focus but touching briefly on the findings in conjunction with Fig.~\ref{fig:STOliveVsPost} we elucidate the impact of dynamical scattering, or, equivalently, specimen thickness, on different signals. A multislice simulation \citep{Rosenauer2007} has been performed for SrTiO$_3$ in $[001]$ projection employing the experimental parameters and using $20\times20$ scan pixels. The data was evaluated for thicknesses of 1\,nm and 30\,nm, addressing both the kinematic case and the situation of elevated thickness in our experiment. The results have been compiled in Fig.~\ref{fig:SimulationSTO}.

Figure~\ref{fig:SimulationSTO}\,a shows the phase grating used in the multislice simulation for structural reference. In figure part~(b) we added the theoretical result that would be obtained for SSB ptychography in case all methodological premises were fulfilled in practice. That is, we generated a 4D-STEM data set by means of the weak phase approximation in eq.~(\ref{eq_wpoa}) for a single slice with the thickness of one SrTiO$_3$ unit cell and performed the SSB reconstruction. Note that this is identical to the phase $\Phi$ in eq.~(\ref{eq_wpoa}), low-pass filtered with a circular aperture that has twice the radius of the probe-forming aperture.

Figures~\ref{fig:SimulationSTO}\,c-j show the results of evaluating the 1\,nm (left column) and the 30\,nm data (right column) using different methods aligned row-wise. As can be expected from former studies employing first moment based imaging \citep{Mueller-Caspary2017,Muller2014b} of centrosymmetric structures, Figs.~\ref{fig:SimulationSTO}\,c-f resemble the atomic structure in terms of momentum transfer vector maps and their divergences with the atoms being sites of central fields. This is preserved in qualitative manner only for more elevated thicknesses. Note that Figs.~\ref{fig:SimulationSTO}\,c and~e are proportional to the probe-convoluted distributions of the projected electric field and charge density, respectively.

Similarly, the SSB reconstructions in Figs.~\ref{fig:SimulationSTO}\,g and~h yield reliable structural contrast at both low and elevated thickness despite the violation of the weak phase approximation in eq.~(\ref{eq_wpoa}), which confirms our interpretation in the post processing section. However, already the result in Fig.~\ref{fig:SimulationSTO}\,g should be considered as qualitative except for the oxygen sites. Please note that the probe had been focused on the specimen surface in the simulation. Because the SSB reconstruction considers the 30\,nm thick specimen as a single slice here, the optimum focus would have been at some depth inside the specimen which is one reason why atomic sites appear slightly broader (Fig.~\ref{fig:SimulationSTO}\,h).

The situation is different for the ePIE results in Figs.~\ref{fig:SimulationSTO}\,i,j. Whereas the reconstruction for the thin specimen in Fig.~\ref{fig:SimulationSTO}\,i represents the phase excellently and can be considered quantitative within the general framework of validity of ePIE, the algorithm has severe difficulties in reconstructing the object transmission function at 30\,nm thickness in Fig.~\ref{fig:SimulationSTO}\,j. In Fig.~\ref{fig:SimulationSTO}\,k the reconstructed probe looks like an airy disc. This is the result of an aberration free probe, limited only by an aperture in diffraction space, and that was used in the simulation. In Fig.~\ref{fig:SimulationSTO}\,l sample information is transferred to the probe. This further confirms our observations concerning ePIE in the post processing section by simulation. However, Fig.~\ref{fig:STOliveVsPost}\,e,h gives even less information from the specimen than Fig.~\ref{fig:SimulationSTO}\,j,l, which can  be attributed to Poisson noise neglected in the simulation, residual aberrations, and importantly, different manners of separating probe and object which starts to fail when dynamical scattering sets in. To conclude, selecting the SSB scheme for live imaging of structural contrast can also be supported from the simulation point of view.

\begin{figure}
    \centering
    \includegraphics[width=\linewidth]{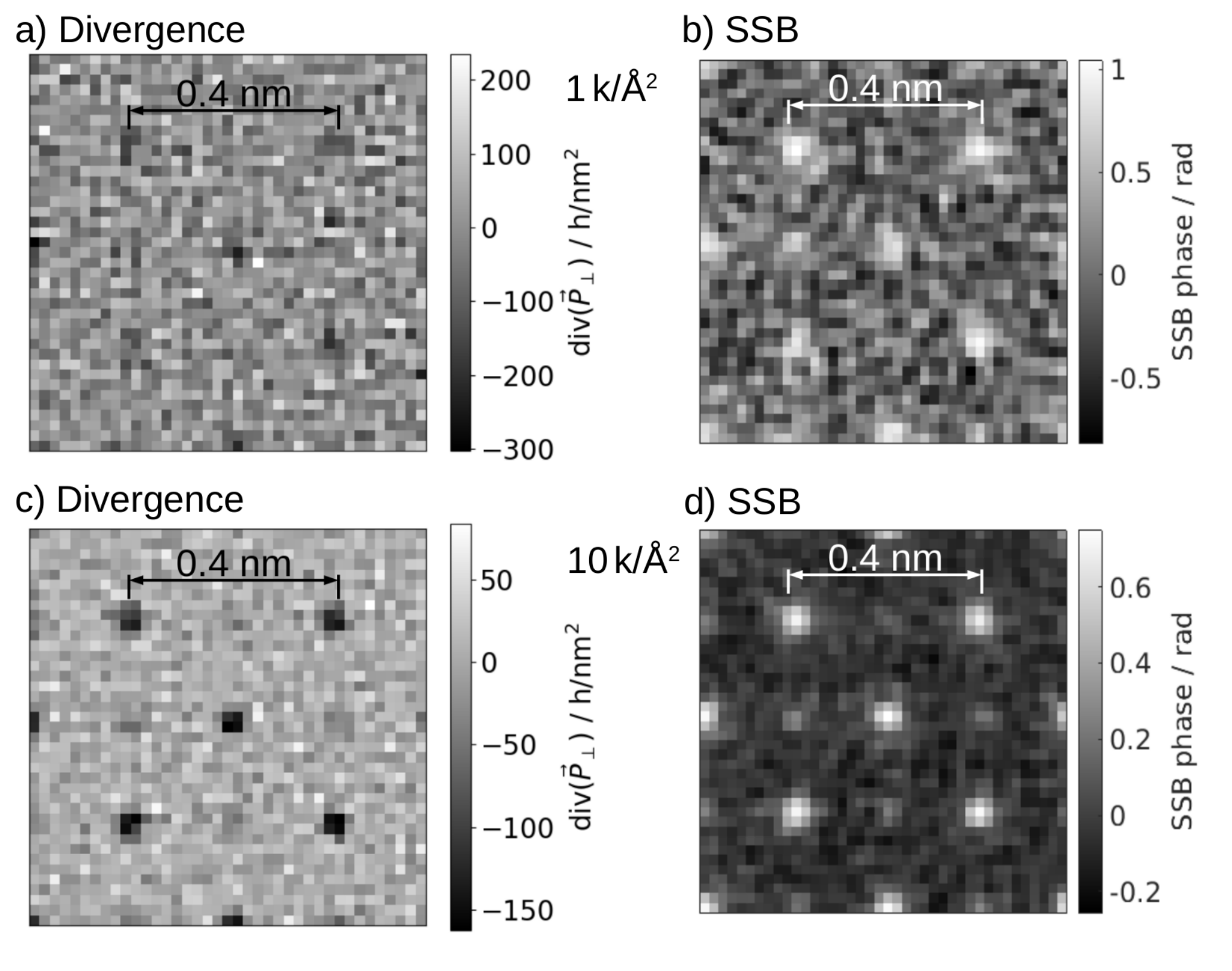}
    \caption{Low dose simulation for SrTiO$_3$: (a, c) Divergence of first moment, (b, d) SSB without using weak phase approximation in the simulation of the 4D-STEM-data, (a, b) at 1000 electrons per \r{A}$^2$, (c, d) at 10000 electrons per \r{A}$^2$. The simulation parameters where the same as in Figure~\ref{fig:SimulationSTO}c.}
    \label{fig:SimulationSTOlowDose}
\end{figure}

\textbf{Low dose.~~}
The performance of the SSB reconstruction and the divergence of the first moment was checked in a low dose simulation~(Fig.~\ref{fig:SimulationSTOlowDose}). At 1000 electrons per \r{A}$^2$~(Fig.~\ref{fig:SimulationSTOlowDose}a,b) the heavy atom columns are visible. At this dose, the SSB reconstruction gives stronger contrast than the divergence of the first moment. At 10000 electrons per \r{A}$^2$~(Fig.~\ref{fig:SimulationSTOlowDose}c,d) also the oxygen atom columns are visible. The SSB reconstruction and the divergence of the first moment shows similar performance at this dose.

\section{Discussion}
The digitisation that took place decades ago with key developments such as charge-coupled device (CCD) cameras and computer controlling, processing and visualisation in STEM denotes one of the drastic paradigm changes in electron microscopy. It enabled the live assessment of recorded data and transformed an optimisation of experimental parameters from multiple sessions to a quick feedback loop taking only several minutes within a single session. Surprisingly, innovative hardware associated with an increase of the dimensionality of the recorded data has, to some extent, put us back to ancient workflows for advanced methodologies. A major challenge for contemporary imaging in the era of Big Data is thus to make current ex-situ multidimensional evaluations capable for live imaging. Within this context, the present work shall be seen as a first step that demonstrates the feasibility of such a workflow using a rather straightforward example. Our work highlights the ongoing push towards high-performance computational methods in electron microscopy that are driven by an increasing camera performance \citep{Weber2020}. That includes suitable software frameworks, connections, storage, processing hardware, and know-how in computer science and engineering to be used in synergy with established and future imaging methodologies.

Several important general conclusions can be drawn from the present study. First, adequate computational hardware is already available for this purpose, given that the mathematical formulation can be adapted to make use of it efficiently. Second, and most importantly, open and well-defined software interfaces which were available for the hardware used here, are key prerequisites to implement nonstandard imaging concepts developed in science into established infrastructures. Third, it can be beneficial to exploit partly simplistic models to achieve live imaging capabilities, exemplified here by the use of the SSB algorithm for a materials science case. On the one hand, it violates inherent assumptions significantly, on the other hand, the qualitative nature of the results is better than one might expect from the weak phase approximation. In particular, the present setup can be considered to be highly beneficial for low-dose ptychographic live imaging of challenging specimen in the fields of structural biology and soft matter in Cryo electron microscopy. A suitable experimental setup with open software interfaces to tap the data stream of a 4D-Cryo-STEM experiment was unfortunately not available to the authors to enable the inclusion of respective examples in the present report. On the other hand, this would not change the methodological setup worked out in this paper using solid-state examples.

The current implementation mainly served to investigate the fundamental feasibility and characteristics of ptychography for live imaging. Integrating the parameter selection and results display in to existing instrument control software could be the next steps to improve the usability, making it more practical to apply routinely in microscopy. In that respect, live focusing, stigmation, and, prospectively, correction of further aberrations based on ptychography are possible. Furthermore, the implementation can be extended to include mitigation of artefacts from the edges that are demonstrated in Figure~\ref{fig:cutout}.

Existing LiberTEM UDFs that were previously only used for offline processing were applied to live data without modification, proving a long-standing design goal of LiberTEM \citep{Clausen2020}. In particular, the UDF interface allowed to disentangle details of the data logistics from the numerical and scientific aspects of the used algorithms. The prototype data decoder and UDF runner could easily keep up with the data rate of the Merlin detector. UDFs with low computational load such as virtual detectors or first moments remained at single-digit CPU load percentages. That means much higher data rates are likely to be possible with a suitable distributed UDF runner implementation. This will be required to support multi-chip cameras such as the Gatan K2 or K3 IS or X-Spectrum Lambda. Computationally intensive operations like ptychography are more challenging to scale to such data rates.

The poor scaling behaviour of the current SSB implementation has proven to be a limiting factor. For illustration, doubling the scan resolution at constant aspect ratio quadruples the number of scan points and results in 16x increased computation time. In future, an implementation that significantly reduces the computational load would be highly desirable to make this technique useful for mainstream data analysis. Ideally there would be a constant memory consumption independent of scan area and $\mathcal{O}\left(N\right)$ or $\mathcal{O}\left(N \log N\right)$ scaling for computation effort as a function of the number of scan points.

\section{Summary}

Live imaging of central 4D-STEM signals such as the ptychographic phase, first moments, their divergence and rotation, as well as flexible virtual detectors has been demonstrated. A direct processing of the data stream of a Medipix3 chip mounted in an aberration-corrected STEM was implemented. An enhanced version is available open-source under \url{https://github.com/LiberTEM/LiberTEM-live}. A prototype was used to generate the live results and is available upon request. The live imaging capability could be demonstrated for two materials science cases In$_2$Se$_3$ and SrTiO$_3$, where single-sideband ptychography proved surprisingly robust against imaging at elevated specimen thicknesses around 20\,nm. It is anticipated that the live imaging approach presented here, and the transfer of such direct workflows to further imaging methods, can also enhance imaging in life sciences where, e.g., ptychography is a promising candidate for high-contrast, low-dose imaging of weakly scattering objects without compromising spatial resolution.

\noindent\small\color{Maroon}\textbf{Acknowledgements }\color{Black}
Knut M\"uller-Caspary, Achim Strauch and Benjamin M\"arz were supported by funding from the Initiative and Network Fund of the Helmholtz Association (Germany) under contract VH-NG-1317 (moreSTEM project). Dieter Weber, Alexander Clausen, Arya Bangun, Benjamin M\"arz and Knut M\"uller-Caspary acknowledge support from Helmholtz within the project "Ptychography~4.0" under contract ZT-I-0025.
Dieter Weber, Alexander Clausen and Rafal Dunin-Borkowski received funding from the European Union’s Horizon 2020 research and innovation programme under grant agreements No. 823717 – ESTEEM3 and No. 780487 - VIDEO.

\normalsize

\section{Supplementary material}
\begin{enumerate}
    \item Video of live view: \url{https://www.youtube.com/watch?v=6s_ewwgOhoI}
    \item Animation of gradual processing:\\ \url{https://ptychography-4-0.github.io/ptychography/algorithms.html}
    \item Full software stack: \cite{clausen_alexander_2020_3982290, libertem-live, ptychography40}
    \item Sample notebooks: https://github.com/Ptychography-4-0/ptychography/tree/master/examples
    \item Selected data sets on Zenodo: \cite{Strauch2021a, Strauch2021}
\end{enumerate}

\bibliographystyle{MandM}
\bibliography{references}
\end{document}